\documentstyle[12pt,epsf]{article}

\begin{document}

\baselineskip=18pt plus 0.2pt minus 0.1pt

\makeatletter

\@addtoreset{equation}{section}
\renewcommand{\theequation}{\thesection.\arabic{equation}}
\renewcommand{\thefootnote}{\fnsymbol{footnote}}
\newcommand{\bm}[1]{\mbox{\boldmath $#1$}}
\newcommand{\calD}{{\cal D}}
\newcommand{\abs}[1]{\left\vert #1\right\vert}
\newcommand{\VEV}[1]{\left\langle #1\right\rangle}
\newcommand{\sgn}{\mathop{\rm sgn}}
\newcommand{\ds}{\displaystyle}
\newcommand{\nn}{\nonumber}
\newcommand{\hanbi}[2]{\frac{\delta #1}{\delta #2}}
\newcommand{\thanbi}[2]{\delta #1/\delta #2}
\newcommand{\cald}{{\cal D}}
\newcommand{\ac}{{\overline{c}}}
\newcommand{\henbi}[2]{\frac{\partial #1}{\partial #2}}
\newcommand{\thenbi}[2]{{\partial #1}/{\partial #2}}
\newcommand{\Half}{\frac{1}{2}}
\newcommand{\CR}[1]{\left[ #1 \right\}}
\newcommand{\wt}[1]{\widetilde{#1}}
\newcommand{\ol}[1]{\overline{#1}}
\newcommand{\G}{\Theta}
\newcommand{\aG}{\overline{\Theta}}
\newcommand{\g}{\theta}
\newcommand{\ag}{\overline{\theta}}
\newcommand{\gM}{m}
\newcommand{\agM}{\overline{m}}

\makeatother

\begin{titlepage}
\title{
\hfill\parbox{4cm}
{\normalsize KUNS-1404\\HE(TH)~96/07\\{\tt hep-th/9608128}}\\
\vspace{1cm}
Causality in Covariant String Field Theory
}
\author{
Hiroyuki Hata\thanks{e-mail address: {\tt
  hata@gauge.scphys.kyoto-u.ac.jp}}
  {}\thanks{Supported in part by Grant-in-Aid for Scientific
  Research from Ministry of Education, Science and Culture
  (\#07640394).} {} and
Hajime Oda\thanks{e-mail address: \tt
  oda@gauge.scphys.kyoto-u.ac.jp}
\\
{\normalsize\em Department of Physics, Kyoto University}\\
{\normalsize\em Kyoto 606-01, Japan}}
\date{\normalsize August, 1996}
\maketitle
\thispagestyle{empty}

\begin{abstract}
\normalsize
Causality is studied in the covariant formulation of free string field
theory (SFT). We find that, though the string field in the covariant
formulation is a functional of the ghost coordinates as well as the
space-time coordinate and the latter contains the time-like
oscillators with negative norm, the condition for the commutator of
two open string fields to vanish is simply given by
$\int_0^\pi d\sigma\left(\Delta  X^\mu(\sigma)\right)^2 >0$,
which is the same condition as in the light-cone gauge SFT.
For closed SFT, the corresponding condition is given in a form
which is manifestly invariant under the rigid shifts of the $\sigma$
parameters of the two string fields.

\end{abstract}

\end{titlepage}

\section{Introduction}
\label{sec:intro}

Study of causal properties of string theory is an interesting
and important subject since string is an extended object and could
behave quite differently from point particle theories.
In fact, Martinec \cite{Martinec} examined causality in free
(non-interacting) open SFT using the light-cone gauge formulation
\cite{Mandelstam,CG,KakuKikkawa}.
He found that the commutator of two string fields with string
coordinates $X^\mu(\sigma)$ and $\wt{X}^\mu(\sigma)$, respectively,
vanishes if the condition
\begin{equation}
\int_0^\pi\! d\sigma\left(
X^\mu(\sigma)-\wt{X}^\mu(\sigma)\right)^2 >0 ,
\label{eq:condition}
\end{equation}
is satisfied.\footnote{
Our flat metric is $g_{\mu\nu}={\rm diag}(-1,1,\cdots,1)$.
Therefore, a space-like vector has positive norm.
}
This condition may look unintuitive since it does not require that
all points on $X^\mu(\sigma)$ are space-like separated from all
points on $\wt{X}^\mu(\sigma)$.
Causality in free light-cone SFT was further studied by Lowe
\cite{Lowe}, and the analysis in the interacting SFT was
carried out in ref.\ \cite{LSU}.

The purpose of this note is to reexamine the causality in free SFT on
the basis of its covariant formulation
\cite{Siegel,FreeSFT}, which is expected to be more suited to the
study of the space-time properties of string theory than
the light-cone gauge formulation.\footnote{
Although ref.\ \cite{Martinec} treats also covariant SFT, our
conclusion is different from that of ref.\ \cite{Martinec}.
}
Besides the choice of the time variable for quantization,
the main differences between the covariant formulation
and the light-cone gauge one are that the string field in the former is
a functional of the ghost string coordinates as well as the space-time
string coordinate $X^\mu(\sigma)$ and that the latter contains
time-like oscillator modes with negative norm.
In spite of these differences, we find that the commutator of two
string fields in the covariant formulation vanishes when the same
condition as (\ref{eq:condition}) is satisfied in the open SFT case.
We also examine the closed SFT, and find that the condition for the
closed string field commutator to vanish is given, instead of
(\ref{eq:condition}), by
\begin{equation}
\min_{0\le\theta\le 2\pi}\int_0^{2\pi}\!d\sigma\left(
X(\sigma+\theta) - \wt{X}(\sigma)\right)^2 > 0 ,
\label{eq:condition_closed}
\end{equation}
which is manifestly invariant under the rigid shift of each
$\sigma$ parameter of $X^\mu(\sigma)$ and $\wt{X}^\mu(\sigma)$.

\section{Commutator of string fields}
\label{sec:commutator}

String field in the covariant formulation of SFT is a functional of
the ghost and the anti-ghost coordinates, $\G(\sigma)$ and
$\aG(\sigma)$, as well as the space-time coordinate $X^\mu(\sigma)$.
In this and the next sections we shall consider free open SFT.
The string coordinates are Fourier-expanded as follows:
\begin{eqnarray}
&&X^\mu(\sigma)=x^\mu+\sum_{n\ge 1} x^\mu_n\cos n\sigma ,\\
&&\G(\sigma)=\g_0 + \sum_{n\ge 1}\g_n\cos n\sigma ,\\
&&\aG(\sigma)= \sum_{n\ge 1}\ag_n\sin n\sigma ,
\end{eqnarray}
where $\G$ and $\aG$ are hermitian.
After fixing the stringy local gauge symmetry by adopting the Siegel
gauge \cite{Siegel} which imposes the condition that the string field be
independent of $\g_0$, the gauge-fixed action of free SFT reads
\begin{equation}
S=-\Half\int {\cal D}Z(\sigma)\,\Phi^\dagger[Z(\sigma)] L\Phi[Z(\sigma)] ,
\label{eq:S}
\end{equation}
where $Z(\sigma)$ denotes the set of string coordinates,
$Z(\sigma)\equiv\left(X^\mu(\sigma),\G'(\sigma),\aG(\sigma)\right)$,
and $L$ is the kinetic operator (prime denotes $\thenbi{}{\sigma}$):
\begin{eqnarray}
&&L=\pi\int_0^\pi d\sigma\left\{
-g^{\mu\nu}\hanbi{}{X^\mu}\hanbi{}{X^\nu} +
g_{\mu\nu}X'^{\mu}X'^{\nu}
-2i\left(\G'\aG + \left(\hanbi{}{\aG}\right)'\hanbi{}{\G}\right)
\right\} \nonumber\\
&&\phantom{L}=
-\left(\thenbi{}{x^\mu}\right)^2 + M^2 ,
\end{eqnarray}
with the $(\mbox{mass})^2$ operator $M^2$ being given in terms of the
Fourier components as
\begin{equation}
M^2=2\sum_{n=1}^\infty\left\{
-g^{\mu\nu}\henbi{}{x_n^\mu}\henbi{}{x_n^\nu}
+\left(\frac{n\pi}{2}\right)^2\! g_{\mu\nu}x_n^\mu x_n^\nu
+2in\left(
-\henbi{}{\ag_n}\henbi{}{\g_n}
+\left(\frac{\pi}{2}\right)^2\!\g_n\ag_n
\right)
\right\} .
\label{eq:M^2}
\end{equation}
The present string field $\Phi$ is Grassmann-even and it is
hermitian,
\begin{equation}
\Phi^\dagger[Z(\sigma)]=\Phi[Z(\sigma)]\ .
\label{eq:Phi_is_hermite}
\end{equation}

Quantization of free open SFT is carried out by taking
the center-of-mass time coordinate $x^0$ as the time variable.
Then, the string field commutator
$\left[\Phi[Z(\sigma)],\Phi[\wt{Z}(\sigma)]\right]$
is derived from the equal-time canonical commutation relation for
$\Phi$ and its conjugate momentum $\Pi=\thanbi{S}{\dot \Phi}=\dot\Phi$
together with the equation of motion $L\Phi=0$.
In order to perform this procedure precisely, we here make the
equivalent calculation based on the local component fields of
$\Phi[Z(\sigma)]$.
Namely, we expand $\Phi$ in terms of the component fields
$\varphi_A(x)$:
\begin{equation}
\Phi=\sum_A \Psi_A(x_n^\mu,\g_n,\ag_n)\,\varphi_A(x) ,
\label{eq:Phi=Psi_varphi}
\end{equation}
where $\Psi_A$ is an eigenfunction of the $(\mbox{mass})^2$ operator
(\ref{eq:M^2}).
The concrete expression of $\Psi_A$ reads
\begin{equation}
\Psi_A=\prod_{n=1}^\infty
i^{M_n^0}h^{(n)}_{M_n^0}(ix_n^0)\cdot
\prod_{i=1}^{D-1}h^{(n)}_{M_n^i}(x_n^i)\cdot
\psi_{\gM_n,\agM_n}(\g_n,\ag_n) ,
\label{eq:Psi_A}
\end{equation}
where the wave function $h^{(n)}_M(x)$ ($M=0,1,2,\cdots$) for the
space-time oscillators
$x^\mu_n$ and the wave function $\psi_{\gM,\agM}(\g,\ag)$
($\gM,\agM=0,1$) for the
ghost oscillators $(\g_n,\ag_n)$ are given respectively as
\begin{equation}
h^{(n)}_{M}(x)=\sqrt{\frac{1}{M!}\sqrt{\frac{n}{2}}}\,
H_{M}(\sqrt{\pi n}x)\,e^{-\pi nx^2/4} ,
\label{eq:h}
\end{equation}
with $H_M$ being the Hermite polynomial, and
\begin{eqnarray}
&&\psi_{0,0}(\g,\ag)=\frac{1}{\sqrt{\pi}}e^{-i\pi\g\ag/2}
=\frac{1}{\sqrt{\pi}}\left(1-\frac{i\pi}{2}\g\ag\right) ,
\nn\\
&&\psi_{1,0}(\g,\ag)=\g e^{-i\pi\g\ag/2}=\g ,
\nn\\
&&\psi_{0,1}(\g,\ag)=\ag e^{-i\pi\g\ag/2}=\ag ,
\label{eq:psi_mm}\\
&&\psi_{1,1}(\g,\ag)=\frac{1}{\sqrt{\pi}}\left(
1+i\pi\g\ag\right)e^{-i\pi\g\ag/2}
=\frac{1}{\sqrt{\pi}}\left(1+\frac{i\pi}{2}\g\ag\right) .
\nn
\end{eqnarray}
Therefore, the index $A$ labeling the oscillator mode is
$A=\{M_n^\mu,\gM_n,\agM_n\}_{n=1,2,3,\cdots}$, and the
$(\mbox{mass})^2$ for the state $A$ is given by
\begin{equation}
M_A^2=2\pi\sum_{n=1}^\infty n\left(
\sum_{\mu=0}^{D-1}M_n^\mu + \gM_n + \agM_n\right)
+ M_0^2 ,
\label{eq:M_A^2}
\end{equation}
where $M_0^2=-2\pi$ is the $(\mbox{mass})^2$ of the lightest (tachyon)
state. Of course, it is meaningless to discuss causality in a theory
containing the tachyon state. However, in the following we pretend that
$M_0^2$ is non-negative since our arguments should apply also to
more realistic tachyon-free superstring field theories.
It should be understood that the product over $n=1,2,\cdots$ in
(\ref{eq:Psi_A}) is ordered in such a way that the wave function with
a larger $n$ is placed at more right-hand position.
Note that each of the factor wave functions in (\ref{eq:Psi_A}),
in particular, $i^{M_n^0}h^{(n)}_{M_n^0}(ix_n^0)$ and
$\psi_{\gM_n,\agM_n}(\g_n,\ag_n)$, are hermitian if we regard
$(x^\mu_n,\g_n,\ag_n)$ as hermitian.

In order to carry out the quantization procedure,
we define the inner-product $\eta_{AB}$ between the wave functions
$\Psi_A$ and $\Psi_B$ by
\begin{equation}
\eta_{AB}=\prod_{n=1}^\infty\left(
\int_{-i\infty}^{i\infty}idx_n^0\cdot
\prod_{i=1}^{D-1}\int_{-\infty}^\infty dx_n^i\cdot
i\!\int\! d\ag_n d\g_n\right)
\Psi_A^\dagger(x_n^\mu,\g_n,\ag_n)
\Psi_B(x_n^\mu,\g_n,\ag_n) ,
\label{eq:inner-prod}
\end{equation}
where the integrations over the Grassmann-odd coordinates $\g_n$
and $\ag_n$ are defined by
$\int\! d\g_n\,\g_n=\int\!d\ag_n\,\ag_n=1$.
The special treatment for the (negative-norm) $x_n^0$ oscillator
should particularly be explained. First, the hermitian-conjugation
$\Psi^\dagger$ should be defined by regarding $x_n^0$ as an hermitian
variable. Second, the integration over $x_n^0$ in
(\ref{eq:inner-prod}) should be performed in the pure-imaginary
direction \cite{AFIO}.
The integration measure $\int{\cal D}Z(\sigma)$ in the action
(\ref{eq:S}) is the same measure as in (\ref{eq:inner-prod}) for the
oscillator modes multiplied by $\int d^Dx$ for the center-of-mass
coordinate $x^\mu$.

The explicit expression of $\eta_{AB}$ for the wave functions
(\ref{eq:Psi_A}) is given by
\begin{equation}
\eta_{A\widetilde{A}}=\prod_n (-)^{M_n^0}
\prod_{\mu=0}^{D-1}\delta_{M_n^\mu,\widetilde{M}_n^\mu}\cdot
\eta_{(\gM_n,\agM_n)(\wt{\gM}_n,\ol{\wt{\gM}}_n)} ,
\label{eq:eta_AA}
\end{equation}
where $\eta_{(\gM_n,\agM_n)(\wt{\gM}_n,\ol{\wt{\gM}}_n)}$
is the inner-product for the ghost oscillator wave functions:
\begin{equation}
\eta_{(\gM,\agM)(\wt{\gM},\ol{\wt{\gM}})}
=i\!\int\! d\ag d\g\, \psi^\dagger_{\gM,\agM}
\psi_{\wt{\gM},\ol{\wt{\gM}}}\  .
\label{eq:eta_mm}
\end{equation}
For our particular choice of $\psi_{\gM,\agM}$ (\ref{eq:psi_mm}), the
non-vanishing component of
$\eta_{(\gM,\agM)(\wt{\gM},\ol{\wt{\gM}})}$
reads
\begin{equation}
\eta_{(0,0)(0,0)}=1 ,\quad
\eta_{(1,0)(0,1)}=-\eta_{(0,1)(1,0)}=i , \quad
\eta_{(1,1)(1,1)}=-1 .
\label{eq:num-eta_mm}
\end{equation}

Then, we shall derive the (anti-)commutation relations between the
component fields $\varphi_A(x)$. Let us define the
sign factor $\beta_A=\pm$ for the wave function $\Psi_A$ by
\begin{equation}
\Psi_A^\dagger = \beta_A \Psi_A .
\label{eq:beta_A}
\end{equation}
The hermiticity (\ref{eq:Phi_is_hermite}) implies that\footnote{
$(-)^A=+$ ($-$) if $\Psi_A$ is Grassmann-even (odd).}
$\varphi_A^\dagger(x)=\beta_A (-)^A \varphi_A(x)$,
and the action (\ref{eq:S}) is expressed in terms of $\varphi_A$ as
\begin{equation}
S=\Half\sum_{A,B}\beta_A(-)^A\eta_{AB}\int\! d^Dx
\,\varphi_A(x)\left(\partial^2 - M_A^2\right)\varphi_B(x) .
\label{eq:S_varphi}
\end{equation}
Imposing the equal center-of-mass time canonical (anti-)commutation
relation,\footnote{
$\CR{\varphi_A, \pi_B}\equiv\varphi_A\pi_B-(-)^{AB}\pi_B\varphi_A$.
}
\begin{equation}
\CR{\varphi_A(x), \pi_B(\wt{x})}_{x^0=\wt{x}^0}
=i\delta_{AB}\delta^{D-1}\left(\bm{x}-\wt{\bm{x}}\right) ,
\label{eq:equal-time-CCR}
\end{equation}
with
$
\pi_A(x)\equiv\thanbi{S}{\dot\varphi_A(x)}=
\beta_A(-)^A\sum_B\eta_{AB}\dot\varphi_B(x)
$,
the (anti-)commutation relation between $\varphi$'s with general
space-time coordinates is given by
\begin{equation}
\CR{\varphi_A(x),\varphi_B(\wt{x})}=
\beta_A(-)^A(\eta^{-1})_{BA}\cdot
i\Delta(x-\wt{x} ; M_A^2) ,
\label{eq:CRvarphivarphi}
\end{equation}
using the invariant function $\Delta$:
\begin{equation}
i\Delta(x ; M^2)=
\int\!\frac{d^Dk}{(2\pi)^{D-1}}\,\sgn(k_0)\,
\delta(k^2+M^2)\,e^{ik\cdot x} .
\label{eq:Delta}
\end{equation}
Having obtained the commutator for the local component fields,
the commutation relation for the original string field $\Phi$
is calculated from eqs.\ (\ref{eq:Phi_is_hermite}),
(\ref{eq:Phi=Psi_varphi}) and (\ref{eq:CRvarphivarphi}) to be given by
\begin{equation}
\left[\Phi[Z],\Phi[\wt{Z}]\right]=
\sum_{AB}\Psi_A^\dagger[Z](\eta^{-1})_{BA}\Psi_B[\wt{Z}]
\cdot i\Delta(x-\wt{x};M_A^2)\ .
\label{eq:CRPhiPhi}
\end{equation}

Our final task in this section is to carry out the summation over the
string modes $A$ and $B$ in eq.\ (\ref{eq:CRPhiPhi}).
For this purpose, we make use of the integral expression of the delta
function,
$\delta(k^2+M^2)=
\int_{-\infty}^\infty d\tau/2\pi
\exp\left\{i\tau\left(k^2 + M^2 \right)
\right\}$,
to rewrite the invariant function (\ref{eq:Delta}) as
a double integral with respect to
$\tau$ and $u=\left(2\tau/x^0\right)k_0$:
\begin{equation}
i\Delta(x ; M^2)=ix^0\int_{-1}^1\! du\int_{-\infty}^\infty\! d\tau
\left(\frac{i}{4\pi\tau}\right)^{(D+1)/2}
\exp\left\{
-\frac{i}{4\tau}\left[\bm{x}^2-(1-u^2)(x^0)^2\right]
+i\tau M^2
\right\} .
\label{eq:Delta2}
\end{equation}
Then, the summation over $A$ and $B$ in eq.\ (\ref{eq:CRPhiPhi}) is
carried out using the formula for $h^{(n)}_M(x)$,
\begin{equation}
\sum_{M=0}^\infty h^{(n)}_M(x) h^{(n)}_M(\wt{x}) e^{i\tau\cdot 2\pi n M}
\!=\!\sqrt{\frac{n}{2\left(1-e^{4\pi i n\tau}\right)}}
\exp\left\{
\frac{\pi n}{4i\sin(2\pi n\tau)}\left[
(x^2+\wt{x}^2)\cos(2\pi n\tau) -2x\wt{x}\right]
\right\},
\label{eq:sumhh}
\end{equation}
and a similar one for $\psi_{\gM,\agM}$,
\begin{eqnarray}
&&\sum_{\gM,\agM}\sum_{\wt{\gM},\ol{\wt{\gM}}}
\psi^\dagger_{\gM,\agM}(\g,\ag)
(\eta^{-1})_{(\wt{\gM},\ol{\wt{\gM}})(\gM,\agM)}
\psi_{\wt{\gM},\ol{\wt{\gM}}}(\wt{\g},\ol{\wt{\g}})
\,e^{i\tau\cdot 2\pi n\left(\gM+\agM\right)}
\nn\\
&&=\psi^\dagger_{0,0}\psi_{0,0}
-i\left(\psi^\dagger_{1,0}\psi_{0,1}
-\psi^\dagger_{0,1}\psi_{1,0}\right)e^{2\pi in\tau}
-\psi^\dagger_{1,1}\psi_{1,1}e^{4\pi in\tau}
\nn\\
&&=\frac{1-e^{4\pi in\tau}}{\pi}\exp\left\{
\frac{\pi}{2\sin(2\pi n\tau)}\left[
(\g\ag + \wt{\g}\ol{\wt{\g}})\cos(2\pi n\tau)
+\g\ol{\wt{\g}}+\wt{\g}\ag\right]
\right\} .
\label{eq:sumpsipsi}
\end{eqnarray}
We obtain the following master expression of the string field commutator:
\begin{eqnarray}
&&\left[\Phi[Z],\Phi[\wt{Z}]\right]=
i\Delta x^0\int_{-1}^1\! du\int_{-\infty}^\infty\! d\tau
\left(\frac{i}{4\pi\tau}\right)^{(D+1)/2}
\nn\\
&&\times\exp\left\{
-\frac{i}{4\tau}\left[(\Delta\bm{x})^2-(1-u^2)(\Delta x^0)^2\right]
+i\tau M_0^2
\right\}
\nn\\
&&\times\prod_{n=1}^\infty
\left(\frac{n}{2}\right)^{D/2}\!\!\frac{1}{
\pi\left(1-e^{4\pi in\tau}\right)^{(D-2)/2}}
\exp\Biggl\{\frac{\pi}{4i\sin(2\pi n\tau)}\biggl(
n\left[\left(x_n^2+\wt{x}_n^2\right)\cos(2\pi n\tau)
-2x_n\cdot\wt{x}_n\right]
\nn\\
&&\qquad
+2i\left[\left(\g_n\ag_n+\wt{\g}_n\ol{\wt{\g}}_n\right)\cos(2\pi n\tau)
+\g_n\ol{\wt{\g}}_n+\wt{\g}_n\ol{\g}_n\right]\biggr)\Biggr\} ,
\label{eq:CRPhiPhi2}
\end{eqnarray}
with $\Delta x^\mu=x^\mu-\wt{x}^\mu$ and $x_n^2=g_{\mu\nu}x_n^\mu x_n^\nu$.

Here, we have derived the string field commutator starting from the
equal-time canonical (anti-)commutation relations for the component
fields. However, looking back the above calculation, the equal-time
canonical commutation relation for the string field $\Phi$ and its
conjugate $\Pi=\dot\Phi$ is seen to be given by
\begin{equation}
\left[\Phi[Z],\dot\Phi[\wt{Z}]\right]_{x^0=\wt{x}^0}=
i\delta^{D-1}(\bm{x}-\bm{\wt{x}})
\prod_{n=1}^\infty\delta(ix_n^0-i\wt{x}_n^0)
\,\delta^{D-1}(\bm{x}_n-\bm{\wt{x}}_n)
\cdot(-i)\delta(\g_n+\wt{\g}_n)\delta(\ag_n+\ol{\wt{\g}}_n) .
\label{eq:CCR_for_Phi}
\end{equation}

\section{Causality in covariant open SFT}
\label{eq:causality}

We would like to argue that the commutator (\ref{eq:CRPhiPhi2})
vanishes if the space-time string coordinates $X^\mu(\sigma)$ and
$\wt{X}^\mu(\sigma)$ satisfy the condition (\ref{eq:condition}).
For this purpose, we need two kinds of regularizations for
eq.\ (\ref{eq:CRPhiPhi2}).
One is for the infinite product over the oscillator mode number $n$ in
eq.\ (\ref{eq:CRPhiPhi2}). As this regularization we shall
adopt either cutting off the mode number $n$ at $n=N$,
or smearing the string field $\Phi$ with smooth test functions
$g^{(n)}(x_n^\mu,\g_n,\ag_n)$ for the higher modes $n\ge N+1$
as was done in ref.\ \cite{Lowe} for the light-cone gauge SFT.
Here we shall first adopt the cutoff method for $n$ and replace the
infinite product $\prod_{n=1}^\infty$ in eq.\ (\ref{eq:CRPhiPhi2})
with $\prod_{n=1}^N$. The other regularization we need is for the
singularities of the $\tau$-integrand of eq.\ (\ref{eq:CRPhiPhi2}),
which we shall explain below.

\begin{figure}[htbp]
\begin{center}
\leavevmode
\epsfxsize=10cm
\epsfbox{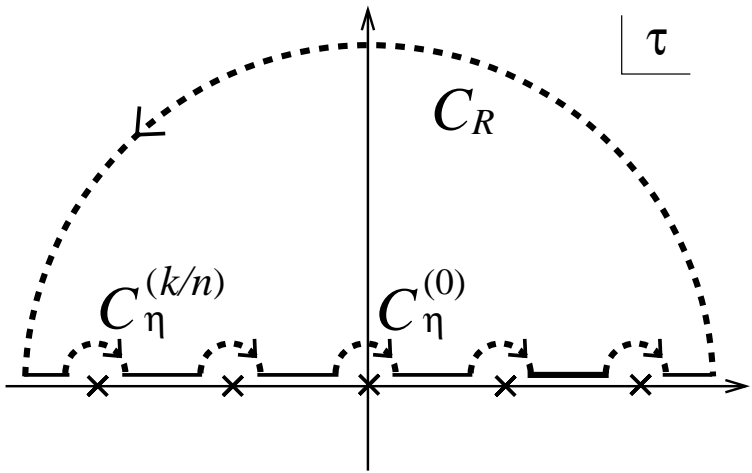}
\caption{
The $\tau$-integration contour. The crosses indicate the
singularities.
}
\label{fig:path}
\end{center}
\end{figure}

Let us consider the $\tau$-integration in eq.\ (\ref{eq:CRPhiPhi2})
along the real axis for a fixed $u$ in the range $|u|\le 1$.
Since the integrand is singular at the origin $\tau=0$ and at
$\tau=\tau_{(k/n)}\equiv{}k/2n$
($n=1,2,\cdots,N$, $k=\pm 1,\pm2,\cdots$), we regularize the
integration by excluding from the integration region
$(-\infty,\infty)$ the neighborhoods of these singular points,
$\abs{\tau}<\eta$ and $\abs{\tau-\tau_{(k/n)}}<\eta$, and then take
the limit $\eta\to 0$.
To evaluate this $\tau$-integration, we consider the contour of Fig.\
\ref{fig:path}, which is obtained by adding small semicircles with
radius $\eta$, $C_\eta^{(0)}$ at the origin and $C_\eta^{(k/n)}$ at
$\tau_{(k/n)}$, and a large one $C_R$ with radius $R$ to the
regularized path (cutoff at $\abs{\tau}=R$) along the real axis.
Since the integrand is non-singular inside the contour, the original
$\tau$-integration of eq.\ (\ref{eq:CRPhiPhi2}) along the real axis
and hence the string field commutator vanish if the contributions from
the semicircles, $C_\eta^{(0)}$, $C_\eta^{(k/n)}$ and $C_R$, vanish in
the limits $\eta\to 0$ and $R\to\infty$.

In order to discuss the conditions for the contributions from
the semicircles to vanish,
we should first understand that the part of the exponent in
(\ref{eq:CRPhiPhi2}) containing the Grassmann-odd variables
$(\g_n,\ag_n)$ and $(\wt{\g}_n,\ol{\wt{\g}}_n)$ has little
effect on the following arguments.
This is because the last exponentiated expression
of eq.\ (\ref{eq:sumpsipsi}) which is apparently singular at
$\tau=\tau_{(k/n)}$ is in fact the sum of three terms
proportional to $e^{2m\pi in\tau}$ ($m=0,1,2$) as the second
expression of  eq.\ (\ref{eq:sumpsipsi}) shows.
Recalling that the product over $n$ in (\ref{eq:CRPhiPhi2}) is now
a finite one, we see that the Grassmann-odd variable parts only affect
the exponent $P$ in eq.\ (\ref{eq:behavior1}) below.

It is easily seen that the contribution from $C_R$ vanishes in the
limit $R\to\infty$ if we have $M_0^2\ge 0$, which we assume true as
explained in the previous section.
Next, let us consider $C_\eta^{(0)}$ at the origin.
Taking the above argument for the terms containing the Grassmann-odd
variables into account, we see that each term of the
$\tau$-integrand in (\ref{eq:CRPhiPhi2}) obtained by expanding with
respect to the Grassmann-odd variables behaves near the origin
$\tau\sim 0$ as
\begin{equation}
\left(\frac{1}{\tau}\right)^{P/2}\exp\left\{
-\frac{i}{4\tau}\left[
(\Delta\bm{x})^2-(1-u^2)(\Delta x^0)^2
+\Half\sum_{n=1}^N\left(\Delta x_n^\mu\right)^2
\right]
\right\} ,
\label{eq:behavior1}
\end{equation}
where we have $\Delta x_n^\mu=x_n^\mu-\wt{x}_n^\mu$, and $P$ is some
integer.
Therefore, the contribution from $C_\eta^{(0)}$ vanishes for every $u$
($|u|\le 1$) if the condition,
\begin{equation}
(\Delta\bm{x})^2-(\Delta x^0)^2
+\Half\sum_{n=1}^N\left(\Delta x_n^\mu\right)^2 > 0 ,
\label{eq:condition0}
\end{equation}
is satisfied.
Taking the limit $N\to\infty$, we obtain the condition
(\ref{eq:condition}).

The estimate of the contribution from $C_\eta^{(k/n)}$ is quite
similar. Here, complication arises due to the fact that different
$(n,k)$'s with coinciding ratio $k/n$ give the same $\tau_{(k/n)}$.
In any case, we find that the contribution from $C_\eta^{(k/n)}$'s
vanish in the limit $\eta\to 0$ if the condition,
\begin{equation}
(x_n\pm \wt{x}_n)^2 \ge 0 ,
\label{eq:condition_n}
\end{equation}
is satisfied for every $n$.
Implication of this condition (\ref{eq:condition_n}) will soon be
explained below.

Although the condition (\ref{eq:condition}) for the covariant SFT is
apparently the same as the one in the light-cone gauge
formulation \cite{Martinec,Lowe}, their expressions in terms of the
oscillator modes $x_n^\mu$ are different: the condition
(\ref{eq:condition}) in the light-cone gauge contains only the
transverse oscillators while (\ref{eq:condition}) for the covariant
formulation depends on $x_n^\mu$ for all $\mu$. In particular,
the time-like oscillator modes $x_n^0$ contribute to
(\ref{eq:condition}) with negative sign;
$-\left(\Delta   x_n^0\right)^2$.
In the light-cone gauge formulation, if the center-of-mass coordinates
of two strings are space-like separated, the condition
(\ref{eq:condition}) is automatically satisfied.
On the other hand, one might think that this property no longer holds
in the covariant formulation due to the
negative sign terms, $-\left(\Delta x_n^0\right)^2$.
However, as explained in Sec.\ 2 the time-like oscillators $x_n^0$
take {\em pure-imaginary} values.
(In the first quantization language, $x_n^0$ is an hermitian operator
but its eigenvalues are pure-imaginary \cite{AFIO}.)
This implies that, in the covariant formulation also, the string field
commutator vanishes if the center-of-mass string coordinates are
space-like separated.
Therefore, although the RHS of eq.\ (\ref{eq:CRPhiPhi2}) is multiplied
by $\Delta x^0$, we do not need to take care of the condition
$\Delta x^0=0$ as another sufficient condition for the string field
commutator to vanish.
The fact that the time-like oscillators $x_n^0$ take pure-imaginary
values also implies that the condition (\ref{eq:condition_n}) is
always satisfied and hence may completely be forgotten.
Summarizing, the commutator of free open string fields vanish if the
condition (\ref{eq:condition}) is satisfied.

The above arguments can be repeated to obtain the same conclusion
if we adopt the smearing regularization \cite{Lowe} instead of
the cutoff one for the mode number $n$.
We do not present the details here but only mention that the argument
of ref.\ \cite{Lowe} applies also to the present ghost coordinate
parts.
The point is that the ghost coordinates part in (\ref{eq:CRPhiPhi2})
reduces to the delta function in the limit $\tau\to 0$:
\begin{equation}
-4in\tau\exp\left\{
\frac{1}{4n\tau}\left(\g_n+\wt{\g}_n\right)
\left(\ag_n + \ol{\wt{\g}}_n\right)\right\}
\to
-i\delta\left(\g_n+\wt{\g}_n\right)
\delta\left(\ag_n + \ol{\wt{\g}}_n\right) .
\label{eq:ghost-delta}
\end{equation}

\section{Causality in closed SFT}
\label{sec:closedSFT}

The arguments of the previous sections for open SFT can be
extended to the closed SFT case. Here we briefly explain the points
particular to closed SFT. The string coordinate
of the closed string is Fourier expanded into both the cosine and the
sine modes as
\begin{equation}
X^\mu(\sigma)=x^\mu+\sum_{n\ge 1}\left(
x^\mu_{Cn}\cos n\sigma + x^\mu_{Sn}\sin n\sigma\right) ,
\label{eq:X_closed}
\end{equation}
for the space-time coordinate $X^\mu(\sigma)$ and similarly for the
ghost coordinates.
An important point for closed SFT is that the string field
$\Phi[Z(\sigma)]$ is subject to the constraint that it be invariant
under the rigid shift of the parameter $\sigma$, namely,
\begin{equation}
\Phi[Z(\sigma+\theta)]=\Phi[Z(\sigma)] ,
\label{eq:cond_on_closed_Phi}
\end{equation}
for an arbitrary $\theta$.
To obtain the correct closed string field commutator which takes into
account the constraint (\ref{eq:cond_on_closed_Phi}), we have to make
the following modifications on eq.\ (\ref{eq:CRPhiPhi2}).
First, noting that, under the shift of $\sigma$ in
(\ref{eq:X_closed}) by an amount $\theta$, the Fourier coefficients
$(x^\mu_{Cn},x^\mu_{Sn})$ is transformed as
\begin{equation}
\pmatrix{x^\mu_{Cn}\cr x^\mu_{Sn}}\to
\pmatrix{x_C(\theta)^\mu_n\cr x_S(\theta)^\mu_n}=
\pmatrix{\cos n\theta & \sin n\theta \cr -\sin n\theta &\cos n\theta}
\pmatrix{x^\mu_{Cn}\cr x^\mu_{Sn}} ,
\label{eq:x_theta}
\end{equation}
we replace the quantity
$\left(x_n^2+\wt{x}_n^2\right)\cos(2\pi n\tau)-2x_n\cdot\wt{x}_n$
in the exponent of eq.\ (\ref{eq:CRPhiPhi2}) by
\begin{equation}
\sum_{\alpha=C,S}\left\{\left(
[x_\alpha(\theta)_n]^2 +[\wt{x}_\alpha(\wt{\theta})_n]^2 \right)
\cos(2\pi n\tau)
-2x_\alpha(\theta)_n\cdot\wt{x}_\alpha(\wt{\theta})_n
\right\} .
\label{eq:exponent_closed}
\end{equation}
A similar replacement is necessary also for the ghost coordinates.
Then, we have to carry out the integrations over $\theta$ and
$\wt{\theta}$:
$\int_0^{2\pi}d\theta/2\pi\int_0^{2\pi}d\wt{\theta}/2\pi$,
which effects the projection of the string field into the subspace
satisfying the condition (\ref{eq:cond_on_closed_Phi}).
Repeating the arguments of the previous section, we see that
the (sufficient) condition for the closed string field
commutator $\left[\Phi[Z],\Phi[\wt{Z}]\right]$ to vanish is that
the inequality,
\begin{equation}
(\Delta\bm{x})^2-(1-u^2)(\Delta x^0)^2 +
\Half\sum_{n=1}^N\sum_{\alpha=C,S}\left(
[x_\alpha(\theta)_n]^2 +[\wt{x}_\alpha(\wt{\theta})_n]^2
-2x_\alpha(\theta)_n\cdot\wt{x}_\alpha(\wt{\theta})_n
\right) >0 ,
\label{eq:cond-pre_closed}
\end{equation}
holds for an arbitrary $u\in [-1,1]$ and arbitrary
$\theta,\wt{\theta}\in [0,2\pi]$.
This implies that the condition (\ref{eq:condition}) should be replaced
for closed SFT by (\ref{eq:condition_closed}).
The condition (\ref{eq:condition_closed}) is indeed invariant under
independent shifts of the $\sigma$ parameters of the two string
coordinates.

\section{Summary}
\label{sec:summary}

In this paper, we have analyzed causality in free covariant SFT.
We obtained, as a sufficient condition for the string field commutator
to vanish, (\ref{eq:condition}) for open SFT and
(\ref{eq:condition_closed}) for closed SFT.
Compared with the light-cone gauge SFT, the string field in covariant
SFT depends additionally on the ghost coordinates and the time-like
oscillator modes of $X^0(\sigma)$.
We found that the ghost coordinates do not enter the analysis of
causality. As for the time-like oscillator modes, the fact that
$x_n^0$ take pure-imaginary values was crucial for the commutability
condition to be given simply by (\ref{eq:condition}) or
(\ref{eq:condition_closed}).

{}From the view point of the component fields, the string field
commutator should vanish if the center-of-mass string coordinates are
space-like separated.
The conditions (\ref{eq:condition}) and (\ref{eq:condition_closed})
imply that the commutative region is ``enlarged'' as a result of the
string mode summation: the commutator vanishes if the center-of-mass
coordinates are space-like separated, however the commutator can
vanish even if the separation of center-of-mass coordinates is
time-like due to the oscillator mode terms in (\ref{eq:condition}) and
(\ref{eq:condition_closed}).

\vspace{.7cm}
\noindent
{\Large\bf Acknowledgments}\\[.2cm]
We would like to thank S.\ Yahikozawa for valuable discussions.

\newcommand{\J}[4]{{\sl #1} {\bf #2} (19#4) #3}
\newcommand{\andJ}[3]{{\bf #1} (19#3) #2}
\newcommand{\MPL}{Mod.\ Phys.\ Lett.}
\newcommand{\NP}{Nucl.\ Phys.}
\newcommand{\PL}{Phys.\ Lett.}
\newcommand{\PR}{Phys.\ Rev.}
\newcommand{\PRL}{Phys.\ Rev.\ Lett.}
\newcommand{\PTP}{Prog.\ Theor.\ Phys.}


\begin{thebibliography}{99}

\bibitem{Martinec} E.\ Martinec,
\J{Class.\ Quant.\ Grav.}{10}{L187}{93}.

\bibitem{Mandelstam}S.\ Mandelstam, \J{\NP}{B64}{205}{73};
\andJ{B83}{413}{74}.

\bibitem{CG}E.\ Cremmer and J.-L.\ Gervais,
\J{\NP}{B76}{209}{74}; \andJ{B90}{410}{75}.

\bibitem{KakuKikkawa} M.\ Kaku and K.\ Kikkawa,
\J{\PR}{D10}{1110}{74}; \andJ{D10}{1823}{74}.

\bibitem{Lowe}D.\ A.\ Lowe, \J{\PL}{326B}{223}{94}.

\bibitem{LSU}D.\ A.\ Lowe, L.\ Susskind and J.\ Uglum,
\J{\PL}{327B}{226}{94}.

\bibitem{Siegel} W.\ Siegel,
\J{\PL}{149B}{157}{84}; \andJ{149B}{162}{84};
\andJ{151B}{391}{85}; \andJ{151B}{396}{85}.

\bibitem{FreeSFT} W.\  Siegel and B.\ Zwiebach,
 \J{\NP}{B263}{105}{86};
K.\ Itoh, T.\ Kugo, H.\ Kunitomo and H.\ Ooguri,
\J{\PTP}{75}{162}{86};
T.\ Banks and M.\ E.\ Peskin,
\J{\NP}{B264}{513}{86}.

\bibitem{AFIO} H.\ Arisue, T.\ Fujiwara, T.\ Inoue, and K.\ Ogawa,
\J{J.\ Math.\ Phys.}{22}{2055}{81}.

\end{thebibliography}
\end{document}